\documentclass{article}
\usepackage{arxiv}
\usepackage[utf8]{inputenc} 
\usepackage[T1]{fontenc}    
\usepackage{xr}
\usepackage{xr-hyper}
\usepackage{hyperref}       
\usepackage{url}            
\usepackage{etoolbox}
\usepackage{multirow}
\usepackage{multicol}
\usepackage{booktabs}       
\usepackage{amsfonts}       
\usepackage{nicefrac}       
\usepackage{microtype}      
\usepackage{graphicx}
\usepackage{doi}
\usepackage[ruled,vlined]{algorithm2e}
\usepackage{amssymb,amsbsy,amsmath}
\usepackage{dcolumn}        
\usepackage[normalem]{ulem} 
\usepackage{nameref}
\usepackage[sort&compress]{natbib}
\usepackage{natmove}
\setcitestyle{super,square,citesep={,}, notesep={;}}
\usepackage{longtable}
\usepackage{zref-xr}
\zxrsetup{toltxlabel}
\usepackage{comment}
\usepackage{subfiles}

\title{City-wide modeling of Vehicle-to-Grid Economics to Understand Effects of Battery Performance}
\author{
	\href{https://orcid.org/0000-0002-9465-3840}{\includegraphics[scale=0.06]{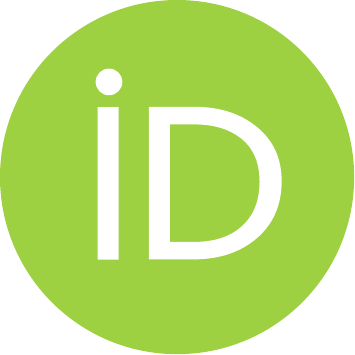}\hspace{1mm}Heta A.~Gandhi}\\
	Department of Chemical Engineering\\
	University of Rochester\\
	Rochester, NY, 14627 \\
	\texttt{hgandhi@ur.rochester.edu} \\
	\And
	\href{https://orcid.org/0000-0002-6647-3965}{\includegraphics[scale=0.06]{orcid.pdf}\hspace{1mm}Andrew D.~White}\thanks{Corresponding Author} \\
	Department of Chemical Engineering\\
	University of Rochester\\
	Rochester, NY, 14627 \\
	\texttt{andrew.white@rochester.edu} \\
}
\date{}

\begin{document}
\maketitle


\begin{abstract}
Vehicle-to-grid (V2G) is a promising approach to solve the problem of grid-level intermittent supply and demand mismatch, caused due to renewable energy resources, because it uses the existing resource of electric vehicle (EV) batteries as the energy storage medium. EV battery design together with an impetus on profitability for participating EV owners is pivotal for V2G success. To better understand what battery device parameters are most important for V2G adoption, we model the economics of V2G process under realistic conditions. Most previous studies that perform V2G economic analysis, assume ideal driving conditions, use linear battery degradation models, or only consider V2G for ancillary services. Our model accounts realistic battery degradation, empirical charging efficiencies, for randomness in commute behavior, and historic hourly electricity prices in six cities in the United States. We model user behavior with Bayesian optimization to provide a best-case scenario for V2G. Across all cities, we find that charging rate and efficiency are the most important factors that determine EV users' profits. Surprisingly, EV battery cost and thus degradation due to cycling has little effect. These findings should help focus research on figures of merit that better reflect real usage of batteries in a V2G economy.\\
\end{abstract}

{\bf Keywords: } Vehicle-to-grid economics;
Electric vehicles;
Stochastic modeling;
Cost-benefit analysis

\section{Introduction}

\begin{figure}[!htb]
    \centering
    \includegraphics[width=0.9\textwidth]{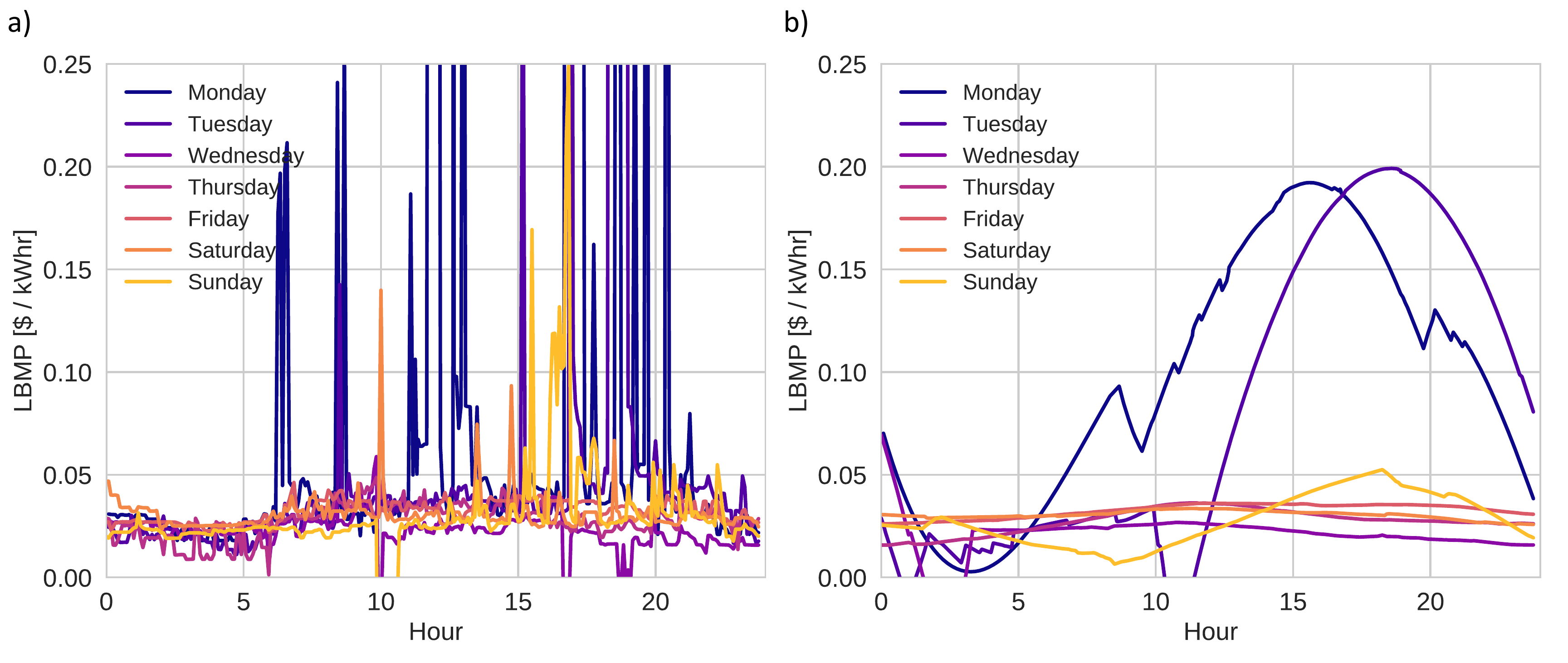}
    \caption{New York City LBMP from October 1, 2018 (Monday) to October 7, 2018 (Sunday). Panel (a) shows the variations induced by RES in the price of electricity. Panel (b), a smoother version of (a), shows the structure of electricity price which is variable due to intermittent nature of RES.}
    \label{fig:lbmp-day-combined}
\end{figure}

Over the past two decades, governments of 37 states, Washington D.C. and 4 territories across the United States have either put forward renewable portfolio standards or set renewable energy targets to address global warming.\cite{portfolio} As a result, the contribution to energy production through renewable energy sources (RES) has increased and RES contributed 17.6\% of the total electricity production in the United States in 2018\cite{nrel}. The intermittent nature of RES causes sudden changes in power availability which leads to demand and supply mismatch and thus electricity price fluctuations\cite{DAINA2017}. Figure~\ref{fig:lbmp-day-combined} shows the variation in location based marginal price (LBMP) in New York City (NYC) for one week. LBMP is defined as the price of producing an additional MW of energy at a given time in a specific location\cite{nyiso}. In this paper, `electricity price' and `LBMP' are used synonymously. As seen in  Figure~\ref{fig:lbmp-day-combined}a, the LBMP varies with the time of the day and it can be very low at some times of the day, even negative and as high as \$1.2/kWh at some other times (not seen in Figure~\ref{fig:lbmp-day-combined}). This variation is more clearly seen in Figure~\ref{fig:lbmp-day-combined}b for Monday and Wednesday where the LBMP varies greatly throughout the day. Figure~\ref{fig:lbmp-day-combined}b is obtained by smoothing Figure~\ref{fig:lbmp-day-combined}a using the Savitzky Golay filter, which removes high frequency noise from the data while preserving the original shape of the data\cite{Savitzky1951}. Smoothing the LBMP data makes it easy to observe the general structure of electricity prices on different days. It is seen that the price of electricity is, in general, higher on some days than the others.

Grid-scale energy storage is a suggested solution for electricity price management as it suppresses power output fluctuations and helps with peak load shaving\cite{Castillo2014, Zame2018}. However, there are challenges in the deployment of large scale energy storage systems. These include technical challenges in integration of the storage facilities with the existing infrastructure, operational safety concerns associated with large scale storage facilities, and the high cost of storage systems.\cite{IRENA2017, USGAO2018, Sharma2019}. Vehicle-to-grid(V2G)\cite{Kempton2005, MULLAN2012} is a strategy studied here for distributed energy storage to supplement grid-scale storage to combat the problem of supply and demand mismatch.\cite{Wang2013} In V2G, participants use the electric vehicle (EV) battery to store energy during times of low demand and sell electricity back to the grid during times of high demand. With increased automobile electrification\cite{McKinsey} and increasing EV battery capacity\cite{EV_growth}, V2G is now possible.

\begin{figure} [!htb]
    \centering
    \includegraphics[width=\linewidth]{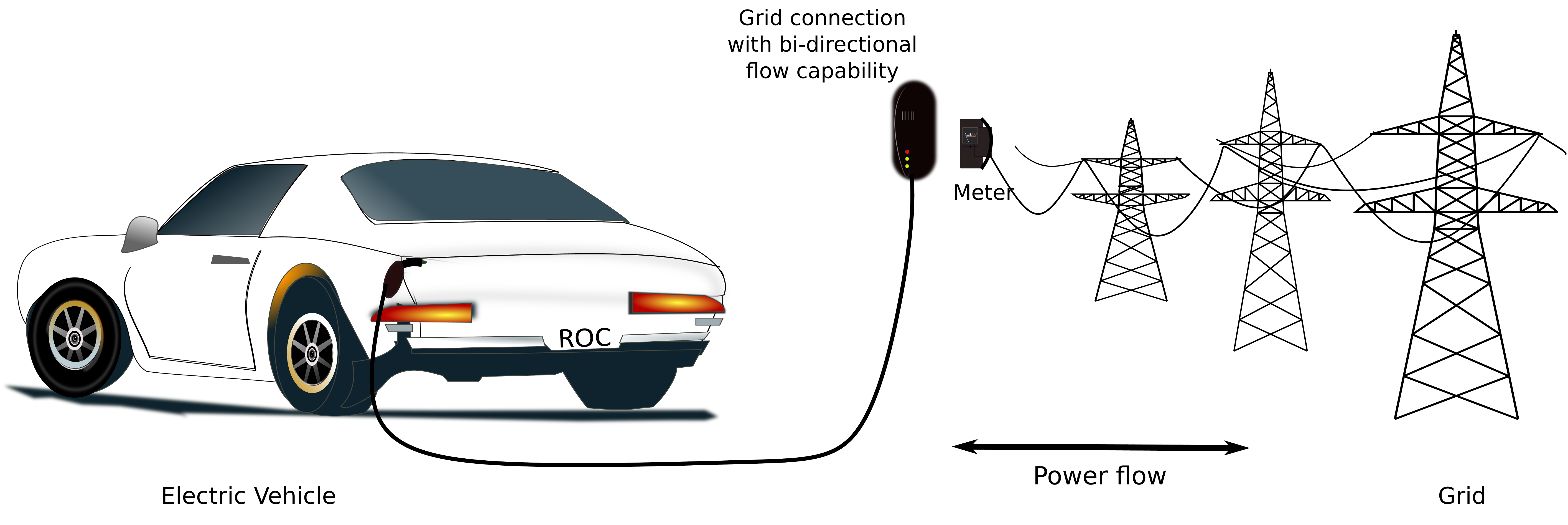}
    \caption{Vehicle-to-grid is bidirectional energy exchange between the grid and electric vehicle. It is a strategy suggested for distributed energy storage.}
    \label{fig:v2g_concept}
\end{figure}

The procedure for a typical V2G operation is characterized by the following: (1) the EV battery is charged during off-peak hours, typically in the middle of the night; (2) the user commutes to work and uses a fraction of the battery capacity while doing so; (3) while the user is at work, the idle car is plugged into the grid at peak hours and electricity is sold to the grid based on whether the grid operator needs power; (4) some battery capacity is reserved to travel back home. After reaching home, the user repeats this cycle. Figure~\ref{fig:v2g_concept} shows a schematic of this operation.

The focus of this article is to assess the effect of battery performance metrics on the microeconomics of V2G implementation in the United States considering the randomness in EV owners' driving patterns and realistic battery degradation. V2G needs infrastructure, like bidirectional energy and data flow between the electricity grid and the EVs, and vehicular design that allows availability of controls and metering on-board\cite{FREEMAN2017, Kempton2005}. Assuming that the infrastructure needed for V2G is already present, will a profit be created for EV owners' if they participate in this program? What are the factors that will affect the savings they earn? How does V2G affect EV battery degradation? How does the charging/discharging rate and efficiency of the EV battery affect profits earned? Do commute patterns of EV users have any affect on V2G economics? A model is presented in this paper to answer these questions. Bayesian optimization is used to find the optimum conditions that will maximize EV owners' benefits. The benefits for EV users are defined as the amount saved on electric fuel for mobility by participating in V2G.

Past research of V2G has studied technical, economical and commercial aspects.\cite{Kempton2005, Kempton2005a, SOVACOOL2009, GUILLE2009, Sortomme2011, ehsani2012, Liu2013, TAN2016, Sovacool2018} Overviews of challenges and benefits of V2G commercialization are presented by \citet{SOVACOOL2009, MULLAN2012} and \citet{HAIDAR2014} 
\citet{PARSONS2014} and \citet{Sovacool2018} present a systematic review of social challenges in V2G implementation. The work of \citet{Kuang2017, Liu2015} and \citet{Maigha2017} focuses on V2G scheduling and optimization. \citet{Kuang2017} proposed an optimization model to study energy sharing between buildings and EV charging stations and the impact of driver behavior on V2G economic performance. \citet{Liu2015} optimized dynamic dispatch of V2G microgrid systems using controlled charging and discharging while \citet{Maigha2017} proposed a day-ahead EV scheduling scheme to optimize V2G energy dispatch. An extensive review of different optimization techniques used in literature is presented by \citet{TAN2016}
V2G impact on the grid is examined in \citet{ GUILLE2009, DRUITT2012} and \citet{Chukwu2017}. \citet{GUILLE2009} and \citet{Chukwu2017} both used aggregated vehicles for energy discharge and studied grid impact but \citet{GUILLE2009} focused more on the effectiveness of vehicle grid integration, while \citet{Chukwu2017} used mathematical modeling of V2G to quantify the amount of energy injected into the grid. The model by \citet{DRUITT2012} simulated EVs to provide demand management and grid balancing services in the UK. Additionally, V2G implementation studies have been conducted in many locations to see how V2G would affect EV owners and electric operators. A summary of some of these studies is given in Table~\ref{tab:literature}.

\newcommand{\colwidth}{0.4}
\newcommand{\smallcol}{0.15}
\begin{table}
    \centering
    \resizebox{\textwidth}{!}{
    \begin{tabular}{m{0.15\textwidth}*{2}{m{\colwidth\textwidth}}*{2}{m{\smallcol\textwidth}}}
        \toprule
         \multirow{2}*{\parbox{\colwidth\textwidth}{\textbf{Location}}} & \multirow{2}*{\parbox{\colwidth\textwidth}{\textbf{System Considerations}}} & \multirow{2}*{\parbox{\colwidth\textwidth}{\textbf{Findings}}} & \multirow{2}*{\parbox{\smallcol\textwidth}{\textbf{V2G Conclusion for Customer}}} & 
         \multirow{2}*{\parbox{\smallcol\textwidth}{\textbf{V2G Conclusion for Grid}}}\\\\
         \bottomrule
        {\begin{flushleft}{United Kingdom (UK) \cite{Ma2012, Wang2016, GOUGH2017}}\end{flushleft}} & 
        \begin{itemize}
            \item An aggregated model of EVs that accounts for SOC, time of day, electricity prices and EV charging requirements
            \item V2G scenarios with a decision making strategy for vehicle deployment  
            \item Analysis of benefits to the grid and cost savings to participating EV owners
        \end{itemize}& 
        \begin{itemize}
            \item Minor impact of EVs on the distribution losses and voltage regulation of the grid
            \item Commute cost reduced by half for EV owners
        \end{itemize}&
        Strongly beneficial &
        Weakly disadvantageous\\
        
        \cmidrule(l){2-5}
         & 
         \begin{itemize}
             \item Agent based coordinated dispatch strategy for EVs and renewable generators
             \item Feasibility and stability of strategy tested on UK generic distribution system and EV owners' profits calculated
         \end{itemize}& 
         \begin{itemize}
             \item Average savings of EV owners is \pounds1 per day which is not enough incentive for participation
         \end{itemize}&
         Weakly beneficial&
         \\
        \cmidrule(l){2-5}
         & 
         \begin{itemize}
             \item Economic feasibility study of vehicle-to-building and V2G for `specific' ancillary services
             \item Accounts for vehicle trip data, electricity market pricing and triads and battery degradation
         \end{itemize}& 
         \begin{itemize}
             \item Net present value of \pounds8400 over 10 years is found to be the highest income generated by participation in the wholesale market and the capacity markets.
             \item Major contributing factors are battery degradation, infrastructure costs, and electricity pricing 
         \end{itemize}&
         Strongly beneficial&
         \\
         \bottomrule
         {\begin{flushleft}{Canary Islands, Spain \cite{Colmenar-Santos2017}}\end{flushleft}} & 
         \begin{itemize}
             \item Cost benefit analysis of vehicle-to-home 
             \item Neglect battery degradation and benefits are calculated as savings in household energy bills
         \end{itemize}& 
         \begin{itemize}
             \item Mobility energy cost of vehicle owners is reduced by 50%
             \item Vehicle-to-home indirectly results in valley filling and peak shaving
         \end{itemize}&
         Strongly beneficial&
         Strongly beneficial\\
         \bottomrule
         {\begin{flushleft}{Firenze, Italy \cite{DEGENNARO2015}}\end{flushleft}} & 
         \begin{itemize}
             \item Model to design and size infrastructure for V2G integration using actual driving patterns
             \item Focus on how energy demand changes with changing infrastructure, no economics
         \end{itemize}& 
         \begin{itemize}
             \item V2G integration is capable of reducing 5\% - 50\% of daily electricity demand in some locations 
         \end{itemize}&\
         &
         Strongly beneficial\\
         \bottomrule
         {\begin{flushleft}{Western Australia\cite{MULLAN2012}}\end{flushleft}} & 
         \begin{itemize}
             \item Utility scale economic analysis
             \item V2G with plug-in hybrid EVs (PHEVs) for peak shaving, ancillary services, demand side management
             \item Accounts for costs of infrastructure, and battery degradation
         \end{itemize}& 
         \begin{itemize}
             \item V2G is not an economical option
             \item Battery capital cost needs to significantly reduce for V2G to become economical for both individuals and industry
         \end{itemize}&
         Weakly beneficial&
         Weakly beneficial\\
         \bottomrule
         {\begin{flushleft}{Canada \cite{Liang2013}}\end{flushleft}} & 
         \begin{itemize}
             \item Policy adjustment scheme to optimize energy delivery through PHEV V2G aggregators for peak shaving
             \item Accounts for randomness in vehicle mobility, battery degradation, and fixed time of use (TOU) pricing
         \end{itemize}& 
         \begin{itemize}
             \item A state dependent policy works best for energy cost minimization of participating vehicles
         \end{itemize}&
         Weakly beneficial&
         \\
         \bottomrule
         {\begin{flushleft}{Texas, United States (US) \cite{Sioshansi2010}}\end{flushleft}} & 
         \begin{itemize}
             \item Economic analysis of benefits of V2G with PHEVs for ancillary services to EV owners and the power system
             \item Accounts for battery degradation cost and day ahead electricity pricing
         \end{itemize}& 
         \begin{itemize}
             \item The cost savings to the power system are about $\$$200 per vehicle in the PHEV fleet participating in ancillary service provision
             \item Driving costs of vehicle owners are slightly reduced
             \end{itemize}&
             Weakly beneficial&\
             Strongly beneficial\\
         \bottomrule
         {\begin{flushleft}{New York City, US \cite{FREEMAN2017}}\end{flushleft}} & 
         \begin{itemize}
             \item Deterministic model analyzing economic benefits to an EV participant from V2G under different scenarios 
         \end{itemize}& 
         \begin{itemize}
             \item Maximum annual savings to the EV owner from the profitable scenarios range between $\$$129 and $\$$231
             \end{itemize}&
             Weakly beneficial&\
             \\
         \bottomrule
    \end{tabular}}
    \caption{Comparison of previous location based V2G implementation studies based on the system used and conclusions of the study.}
    \label{tab:literature}
\end{table}

From the Table~\ref{tab:literature}, it is observed that early studies focused on using PHEVs for V2G. PHEVs have much lower energy capacity than EVs and thus, are capable of lower energy storage. Most of these studies consider the cost of battery degradation and electricity pricing data to analyze the economic benefits of V2G to vehicle owners or the power grid. In the work by \citet{Liang2013}, the starting time of commute and commute distance of vehicle owners are randomly sampled but this sampling is not supported by real data. The model by \citet{GOUGH2017} is the only model that considers the stochasticity in driving patterns of EV owners but their economic analysis is for specific market conditions such as ancillary service provision during triads.\footnote{Triads are defined as the three half-hours of the highest demand on the Great Britain electricity transmission system between November and February each year. The price of electricity is on average $\sim \pounds 35.5/kWh$(\$50/kWh).} This cannot be generalized for electricity markets around the world. The aim of this paper is to build a realistic model that does a cost-benefit analysis for V2G participants in different regions of the United States, taking into account historical electricity LBMP, realistic battery degradation, variable driving patterns and work schedules of EV drivers, and different EV models available in the market. A stochastic model for peak load shaving through V2G is presented. A deterministic model is one where results depend on selection of constraints and assumptions. In a stochastic model, randomness is introduced at different points in the model. In our model, electric cars (battery capacity) and user driving patterns and work schedules are randomized. To assess effects of battery performance metrics, we study the effect of charging/discharging rates under a realistic degradation model and extrapolate lithium-ion battery costs. In order to present the optimal scenario under randomness, we use Bayesian optimization, which provides an upper limit of real performance for maximizing benefits of EV users. Our results indicate that benefits are location dependent and users' time of work matters most, among the randomized parameters, in determining the savings. 

\section{Methods}
Our model calculates the profit of participating in the V2G scheme for EV owners in six cities of the United States using historic LBMP data and accounting for battery degradation and variable work and commute patterns of EV users. This stochastic model was constructed in Python\footnote{Python is a high level programming language with various scientific computing packages \cite{Oliphant2015}.} and is designed to answer questions arising from the economics of V2G technology. LBMP varies throughout the day depending on the demand-supply dynamics and hence, the time of charging or discharging the vehicle affects the cost of charging the EV and profit earned from V2G. This makes it necessary to account for commute patterns and work hours of EV users to realistically calculate their economic benefits. There were 5298 EV users in New York City (NYC) as of December, 2019\cite{nyserda} and each of them may have different commute parameters leading to different savings from V2G. For example, savings from V2G may be different for an EV owner who works night shifts and an EV owner who typically works during the day since price of electricity is different during these times. The presented model jointly samples values for commute time and distance for every EV user from the data available from National Highway Travel Survey (NHTS, 2009) \cite{NHTS}. The arrival times to work and hours worked per week are also sampled independently for every user from the American Community Survey data (ACS, 2016)\cite{PUMS}. The LBMP also varies by the day and to account for the impact of this daily variation on the benefits, vacation time of users is randomly sampled to be between 1-3 weeks.
EV sales in United States increased 79\% in 2018 from 2017 \cite{EV_growth} and with increased EV adoption, more automobile manufacturers have launched their electric car models. Thus, the model is designed to account for different EV models. 2019 United States EV sales data is used to sample EV models for V2G participants.

The model does a cost-benefit analysis for V2G participants under two scenarios as described by \citet{FREEMAN2017}
The price-taker scenario assumes that the user sells electricity whenever at work regardless of the cost of electricity at that time. Whereas the optimal selling price (OSP) scenario offers some control to each user by letting them fix a selling price at the start of every year and electricity is sold if the LBMP is greater than that fixed price. So, while at work, the EV is plugged-in to the grid but energy is sold only when LBMP at a given time exceeds the user-defined OSP. In a 'mean model', each user in a particular city would have the same OSP which may lead to losses for some users since the costs incurred by them may exceed the cost of commute. Hence, OSP for each user is optimized such that profits from V2G are maximized. Bayesian optimization\cite{Jones1998} was used to optimize the selling price for each user. LBMP data is obtained from Independent System Operators\cite{nyiso, neiso, caiso, pjm} i.e. the operators for competitive wholesale markets in the different states.

The probability distribution for making a net annual profit of $n$ by participation in V2G by an EV owner can be calculated from the joint probability of $n$ and the vector $\vec{\theta}$ and converting it to a conditional probability as in Equation~(\ref{eq:prob_of_profit}). $\vec{\theta}$ is a vector of the model constants $\vec{\theta_f}$ and random variables $\vec{\theta_r}$. $P(n|\vec{\theta_r}; \vec{\theta_f})$ is the conditional probability of earning a net annual profit $n$ given $\vec{\theta}$. The integral in Equation~(\ref{eq:prob_of_profit}) is over the sample space of $\vec{\theta_r}$. Since, $\vec{\theta_f}$ are fixed parameters, they do not affect $P(n)$. For a particular set of stochastic variables $\vec{\theta_r}$, $P(n|\vec{\theta_r})$ becomes deterministic and is written as $N(\vec{\theta})$. $N(\vec{\theta})$ is given by Equation~(\ref{eq:profit}). 
\begin{equation} \label{eq:prob_of_profit}
    P(n) = \int P(n,\vec{\theta}) d\vec{\theta}\\
         = \int P(n|\vec{\theta_r}; \vec{\theta_f} ) P(\vec{\theta_r}) d\vec{\theta_r} \\
         = \int \delta(N(\vec{\theta}) - n) P(\vec{\theta}) d\vec{\theta_r}
\end{equation}
$\vec{\theta_f}$ and $\vec{\theta_r}$ are defined in  Equations~(\ref{eq:theta_f}) and (\ref{eq:theta_r}), respectively.
\begin{equation} \label{eq:theta_f}
    \vec{\theta_f} = [DoD, \eta, r_{ch}, r_d]\\
\end{equation}
\begin{equation} \label{eq:theta_r}
    \vec{\theta_r}  = [d_{c}, \Delta t_{c}, t_w, \Delta t_w, E_{max}]
\end{equation}
$DoD$ is depth of discharge, $\eta$ is the one way efficiency of charging/discharging, $r_{ch}$ is the rate of charging, $r_d$ is the rate of discharging,  $\Delta t_{c}$ is the commute time, $d_{c}$ is the commute distance of the EV user, $t_w$ is the time of work, $\Delta t_w$ is the number of hours worked by the user, and $E_{max}$ is the battery capacity in kWh depending on the EV model. 

\begin{equation} \label{eq:profit}
    N(\vec{\theta}) = \int \big(\phi(t) \ LBMP(t) - c_{deg}(t)\big)dt
\end{equation}

$N(\vec{\theta})$ is a time integral and it depends on power charged or discharged $\phi$ in $kW$ at a given time $t$, the LBMP at that time in $\$/kWh$, the differential battery degradation cost for the power transacted, $c_{deg}(t)$ in $\$/h$. V2G activity is assumed to be done only on working days and the US Federal Calendar is used to determine the holidays. $\phi(t)$ is defined in Equation~(\ref{eq:phi}) as a combination of Heaviside step functions which determine the power discharged or charged when the EV is plugged in into the grid.

\begin{multline} \label{eq:phi}
    \phi(t) = I_w(t)\ c(t, p)\ \eta\ r_d \bigg[H\bigg(SoC(t) - \big(1 - DoD - \frac{2 E_c}{E_{max}}\big)\bigg) - H\big(SoC(t) - 1)\bigg]\\ + I_h(t)\ \eta\ r_{ch} \bigg[H\big(SoC(t) - 1 + DoD\big) - H\big(SoC(t) - 1\big)\bigg] 
\end{multline}

Where $E_c$ is the energy capacity used to commute. $r_{ch}$ and $r_d$ are the rates of charging and discharging, respectively and are treated equal at 11.5kW assuming that a J1772 level 2 charger is used for all charging and discharging purposes. As determined by \citet{Apostolaki-Iosifidou2018}, a round-trip efficiency of 70\% is used for determination of charging and discharging costs of the EV owners. Hence, $\eta$ is assumed to be 83.7\%, the square root of the round trip efficiency 70\% \cite{Apostolaki-Iosifidou2018}. Discharging the battery to 100\% depth (called deep discharge) accelerates the degradation process and less the $DoD$, the better it is for the battery lifetime\cite{Santhanagopalan2015}. The DoD for this model is fixed to be 90\% so that battery degradation is not accelerated but there is enough battery capacity that can be used for commute and V2G.
State of Charge (SoC) of the battery is an important consideration when charging or discharging because the EV battery must not be overcharged or discharged in excess of what can be sold at any given time. It is calculated as a time integral of $\phi(t)$.

\begin{equation} \label{eq:soc}
    SoC(t) = \int_0^t \phi(t') dt'
\end{equation}

$c(t, p)$ is a control function used to ensure that electricity is sold only if LBMP is greater than the set selling price, $p$, at any given time $t$. 
\begin{equation}
    c(t, p) = H(LBMP(t) - p)
\end{equation}
\begin{equation}
    p = \underset{p^*}{\mathrm{argmax}}\ N(p^*| \vec{\theta})
\end{equation}

$I_w$ and $I_h$ are time dependent indicator functions that indicate the state of the EV user as `at work' or `at home', respectively. The charging or discharging of EVs is also dependent on state of the user at a given time besides the SoC of EV battery at that time. If the EV user is at work, the EV is said to be discharging (selling electricity) and if at home, the EV is said to be charging (buying electricity). 
\begin{equation} \label{eq:at_work}
    I_w(t) = H(t - t_w) - H(t - t_w - \Delta t_w)
\end{equation}
\begin{equation} \label{eq:at_home}
    I_h(t) = H(t - t_h) - H(t - t_w + \Delta t_c)
\end{equation}
\begin{equation}
    t_h = t_w + \Delta t_w + \Delta t_c
\end{equation}

 where $t_w$ is the time of work and is also treated as the start time for discharging, $\Delta t_c$ is the commute time in hours, and $t_h$ is the time when user arrives home from work and is treated as charging start time. $t_w$ is sampled stochastically from data\cite{PUMS} :
\begin{equation} \label{eq:5}
    t_w \sim P(t_w)
\end{equation}
`$\sim$' indicates that samples were drawn probabilistically.
The energy needed for commute, $E_c$ is calculated as
\begin{equation} \label{eq:energy_commute}
    E_c = d_c \frac{E_{max}}{d_r}
\end{equation}

where $d_c$ is the commute distance in miles and $d_r$ is the rated range of the vehicle in miles. $E_{max}$ is sampled stochastically from the 2016 EV sales data and the rated range of an EV is provided by the EPA\cite{USEPA}. 
\begin{equation} \label{eq:capacity}
    E_{max} \sim P(E_{max})
\end{equation}

The NHTS dataset \cite{NHTS} is used to jointly sample the commute time and commute distance for the EV users. 
\begin{equation}
    d_c, \Delta t_c \sim P(d_c, \Delta t_c)
\end{equation}

Finally, the probability of earning savings, S, from V2G is calculated as the probability of difference between the total profit from charging/discharging that the EV owner earns while participating in V2G, $N$, and the total profit the user earns for normal commute without V2G, $N_c$.
\begin{equation} \label{eq:10}
    P(s) = P(n - n_c)
\end{equation}

The profit for commute only scenario $N_c(\vec{\theta})$ is calculated in a similar fashion as V2G except that there is no selling of electricity. The duration of charging differs for commute only and V2G scenario because the energy capacity used throughout the day is different in both scenarios. 

One important factor that contributes to the total cost to EV users is the battery degradation cost, $c_{deg}$. Battery degradation is the gradual loss of capacity over the lifetime of a battery. Accurately modeling the battery degradation is relatively difficult because there are a variety of hypothesized mechanisms that reduce capacities of batteries and each is specific to the cathode, anode, and electrolyte choice\cite{Kabir2017}. An overview of the state-of-the art in lithium ion battery degradation modeling is given in the work by \citet{THOMPSON2018}. The EV battery undergoes wear not only due to additional charge/discharge cycles (cycle aging) but also when the battery is resting (calendar aging)\cite{Keil2016}. Temperature ($T$), SoC, DoD and number of cycles ($N_{cyc}$) are the factors that contribute to battery degradation\cite{Smith2017}. In this work, a modified version of the semiempirical lifetime prediction model developed by National Renewable energy Laboratory (NREL)\cite{Smith2012, Santhanagopalan2015} is used. This model takes into account all the above mentioned factors to estimate the battery capacity fade due to calendar aging as well as cycle aging. Capacity fade due to calendar aging is dominated by solid-electrolyte interphase growth resulting in loss of cyclable lithium while that due to cycle aging is controlled by active material loss and mechanical failure\cite{Smith2012, Santhanagopalan2015}. Equations~(\ref{eq:QLi}) -- (\ref{eq:Qloss}) are the governing equations for capacity fade. 
\begin{equation} \label{eq:QLi}
    Q_{li} = b_0 + b_1 t^{-1/2}
\end{equation}
\begin{equation} \label{eq:Qsites}
    Q_{sites} = c_0 + c_1 N_{cyc}
\end{equation}
\begin{equation} \label{eq:Qloss}
    Q = \min(Q_{li}, Q_{sites})
\end{equation}
Where $Q$ gives the capacity fade at time $t$. $Q_{li}$ and $Q_{sites}$ are the calendar aging and cycle aging components, respectively, of battery capacity fade. Coefficients $b_0, b_1, c_0, c_1$ are degradation rate constants that depend on the aging condition of battery. Details of the model can be found in \citet{Santhanagopalan2015}

The battery degradation cost at a given time is then calculated from capacity fade using Equation~(\ref{eq:deg_cost}).
\begin{equation} \label{eq:deg_cost}
    c_{deg} = \frac{c_b \ E_{max} (1 - Q)}{SF}
\end{equation}
where $c_b$ is the capital cost of the battery in \$/kWh. $E_{max} (1 - Q)$ gives the battery degradation in kWh because $Q$ obtained from Equation~(\ref{eq:Qloss}) is the relative battery capacity left after degradation. We introduce SF which is the saturation factor and is defined as the percentage degradation after which the EV battery needs to be replaced.

The above stochastic model was implemented for commuting EV owners participating in V2G in six US cities with real commuting data. A discussion of results is presented in the next section.

\section{Results and Discussion}
It is found that OSP scenario produces positive savings for the EV users while the price-taker scenario adds to the existing commute cost and makes V2G unprofitable. The variance in the magnitude of savings in both scenarios comes from the randomness introduced in the selection of EV models, and charging and discharging time due to the randomness in commute parameters and work patterns. We find that, in all cities, time of work significantly impacts the magnitude of savings.
The effect of charging rate, and battery efficiency on V2G savings was also studied along with the effect of battery capital cost. While charging rate is found to be a crucial factor in increasing savings, battery capital cost is found to have little effect on V2G savings.

\subsection{V2G Economics in Different Cities}
\begin{figure}
    \centering
    \includegraphics[width=0.6\textwidth]{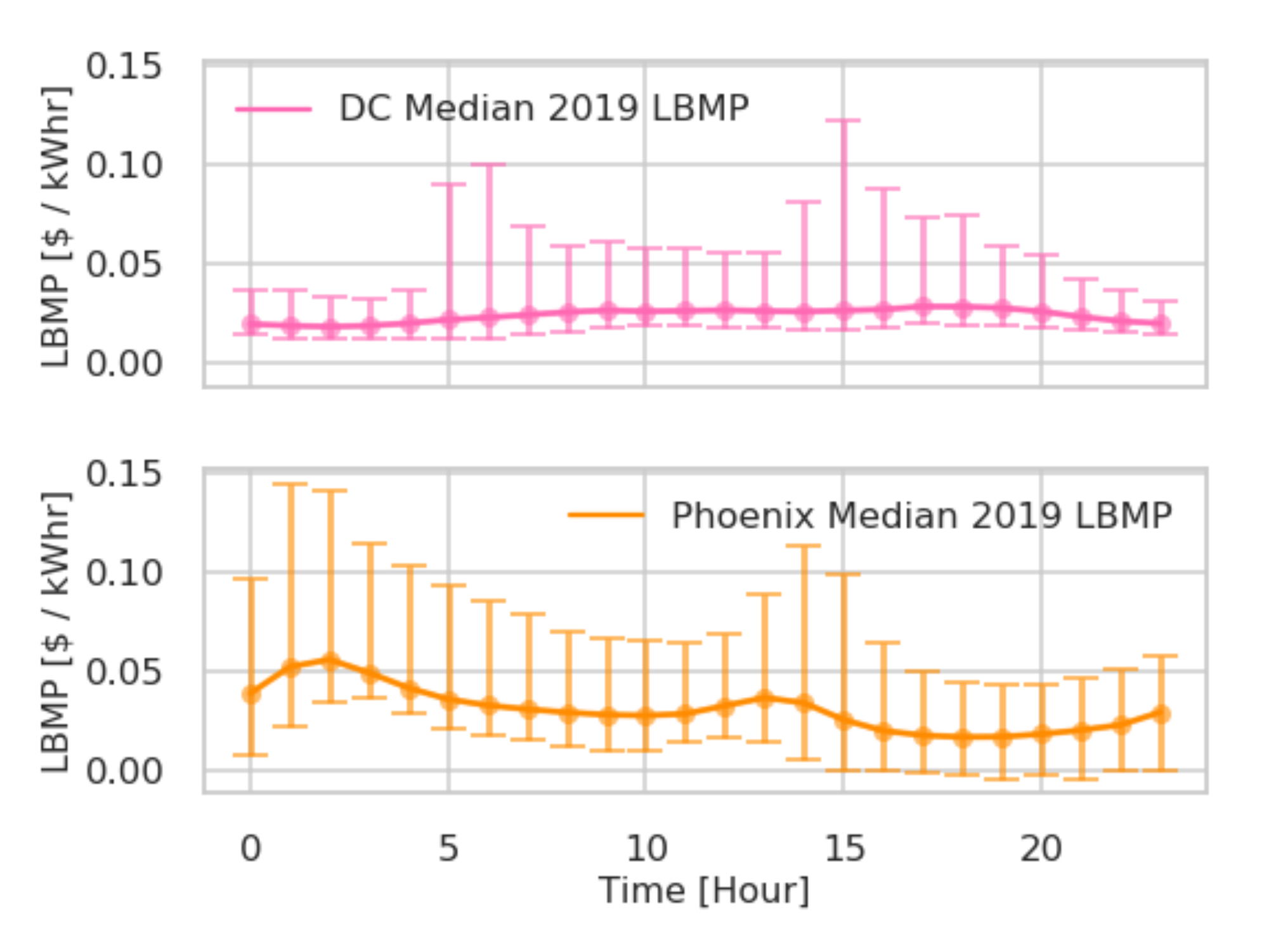}
    \caption{Median annual hourly LBMP for DC and Phoenix. The median electricity prices are somewhat constant throughout the day in DC. In Phoenix, the LBMP is generally higher between 1 pm and 4 pm and again between 12 am and 5 am. This variation in LBMP results in different electricity selling patterns for EV owners in these cities.}
    \label{fig:lbmp_variation}
\end{figure}

Commute patterns and electricity prices change from one city to the other. To see how V2G economics change with changing commute patterns and LBMPs, the economic analysis is applied to the cities of Boston, Chicago, Washington DC (DC), NYC, Phoenix, and San Francisco City (SFC). The commute parameters are sampled for each city using the ACS\cite{PUMS} and NHTS\cite{NHTS} datasets. 2019 LBMP data for respective cities are used to calculate the profitability of V2G. The LBMP for different cities was acquired from the local electric power markets\cite{caiso, neiso, nyiso, pjm}. The electricity pricing data is available on the websites of these electric market operators and the model uses historic TOU LBMP data for all the cities. Figure~\ref{fig:lbmp_variation} shows the difference in LBMP for DC and Phoenix as an example. The LBMP in Phoenix, on average, rises between 1pm and 4pm and again between 12am and 5am. However, in DC, the median LBMP is nearly constant throughout the day. Such differences in LBMP are also observed for other cities.

\begin{figure}
    \centering
    \includegraphics[width = \textwidth]{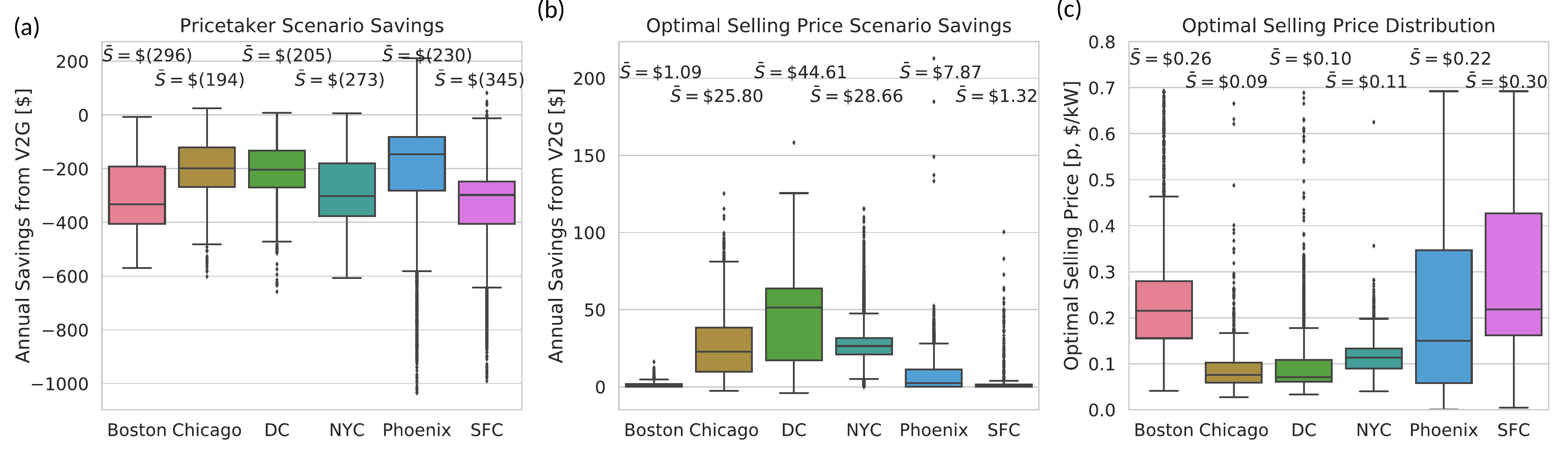}
    \caption{Panel (a): Price-taker scenario results for different cities in the United States. The box plot shows annual savings of the EV users in the cities of Boston, Chicago, DC, NYC, Phoenix, and SFC for the price-taker scenario. $\bar{S}$ indicates the mean annual savings for that city. EV users make negative savings in this scenario i.e. they end up spending more by participating in V2G. Panel (b): OSP scenario results for different cities in the United States. These are annual savings of the EV users for the OSP scenario where every user sets his own selling price. The difference in the distribution of annual savings in the cities is due to location based differences in the distribution of commute parameters and work patterns. Panel (c): Distribution of OSP, p, for different users in the cities. OSP distribution is dependent mainly on the average LBMP prices in the city. In general, EV users working fewer than 5 hours per week have p = \$0/kWh. }
    \label{fig:pt_osp}
\end{figure}

\begin{figure}
    \centering
    \includegraphics[width=0.8\textwidth]{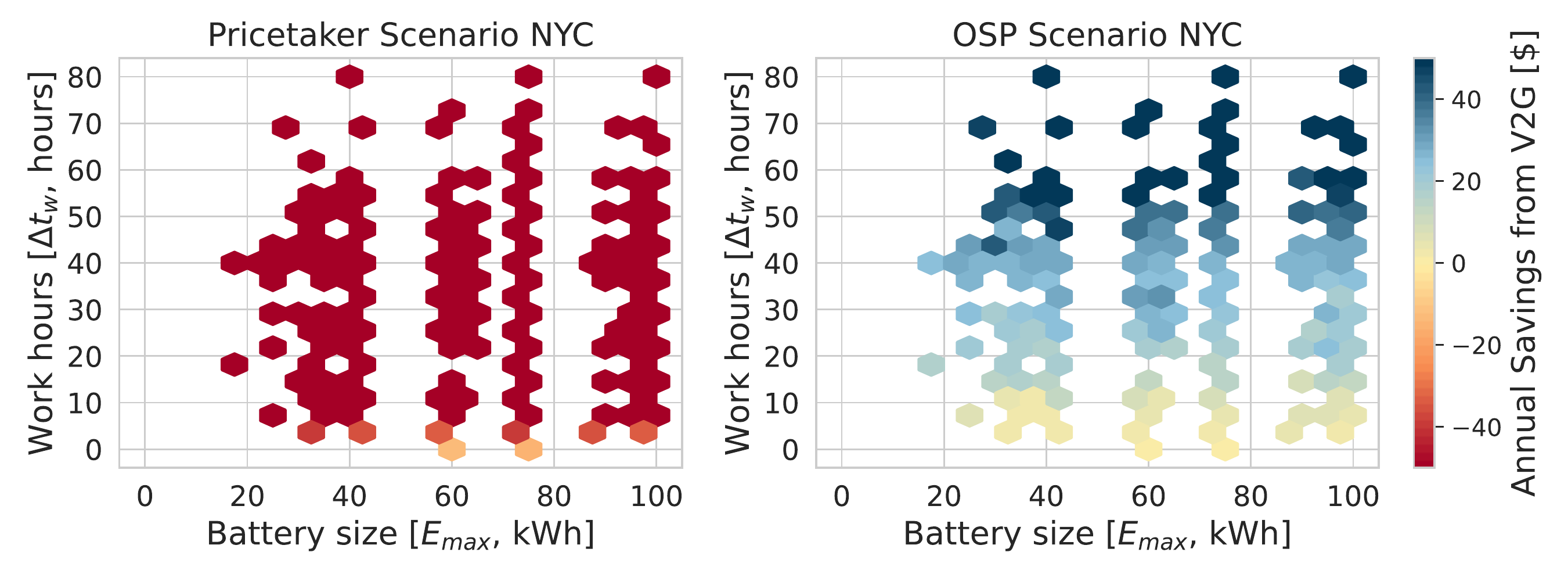}
    \caption{Annual savings from V2G in the Price-taker and OSP scenario as a function of battery size and average number of hours worked per week in NYC. The yellow region indicates users who make close to \$0 savings. The users who work fewer than 10 hours per week save very little by participating in V2G while battery size doesn't have a significant impact on V2G savings. For other cities under consideration, a similar pattern is observed.}
    \label{fig:corr_battery_hours_nyc}
\end{figure}

The results in Figure~\ref{fig:pt_osp} illustrate economics of V2G in different cities. There is no profit in the price-taker scenario and the users spend more by participating in V2G. This is seen in Figure~\ref{fig:pt_osp}a. In the OSP scenario, users in all cities have positive savings because the OSP, $p$, is optimized so that the total cost to the users is less than the cost borne for commute without V2G (Figure~\ref{fig:pt_osp}b). The maximum possible annual savings vary by city. Figure~\ref{fig:pt_osp}c shows the distribution of OSP, $p$, for the six cities in consideration. The general structure of LBMP throughout the day and the number of work hours per week play a major role in determining the average distribution of OSP. It is found that EV owners having who work fewer than 5 hours per week have osp of $\sim$\$0/kWh and those working fewer than 10 hours have close to zero savings. The correlation between savings and number of working hours is illustrated in Figure~\ref{fig:corr_battery_hours_nyc} for NYC. When the number of working hours is small, the window for selling for V2G is short and results in smaller or no savings. In general, potential for savings increases with the number of weekly working hours. On the other hand, battery size has little effect on savings. Figure~\ref{fig:battery_workhours_all} shows this variation for the other cities. Boston and SFC have low savings but a similar correlation with the number of working hours is observed.

\begin{figure}
    \centering
    \includegraphics[width=0.99\textwidth]{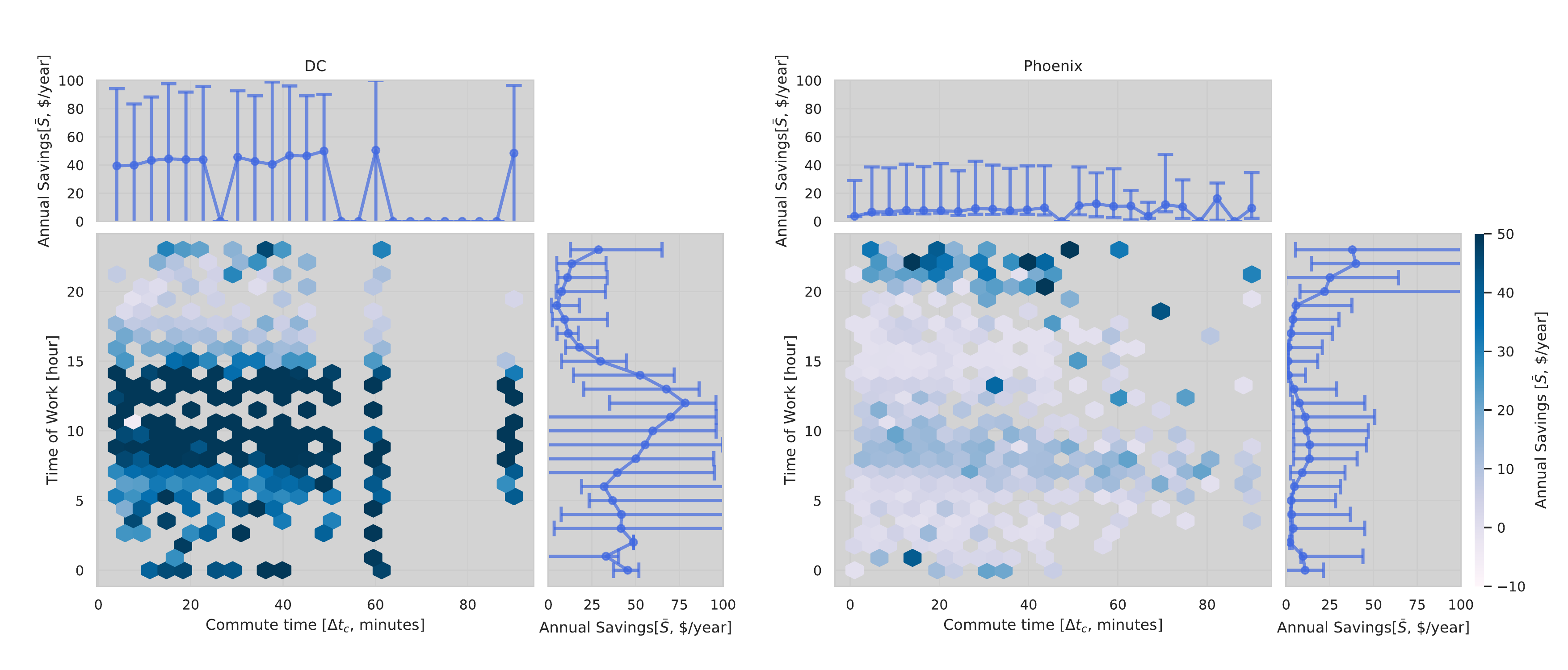}
    \caption{Variation in annual savings with commute time and time of work in DC and Phoenix for the OSP scenario. For each city, the depth of the color in a hexagonal bin indicates average annual savings for that commute time and time of work in that city. In Phoenix, annual savings are high between 8 pm and 2 am and the time of work has a significant effect on the annual savings while in DC, annual savings are generally high irrespective of time of work. There is little effect of commute time on savings. }
    \label{fig:time_DC_AZ}
\end{figure}

The variation in savings in different cities comes from the fact that commute patterns and work patterns differ in these cities. Profit favouring time of work is significant for high annual savings besides the average weekly working hours. In Figure~\ref{fig:lbmp_variation}, it is seen that the LBMP in Phoenix is low during most hours except early afternoon and late night. Hence, the EV users going to work during those peak hours are likely to make bigger savings while EV users going to work other than those peak hours have smaller savings in Phoenix. This can further be seen in Figure~\ref{fig:time_DC_AZ} which compares the effect of commute time and time of work on the annual savings in DC and Phoenix. Notice that the EV users in Phoenix make greater savings if they start work between 8 pm and 2 am. The commute time doesn't seem to have as much effect on the annual savings as the time of work. Savings in DC are highest for start time of work between 8 am and 3 pm and very low between 5 pm and 11 pm which is also consistent with the LBMP pattern for DC seen in Figure~\ref{fig:lbmp_variation}. Figure~\ref{fig:time_all} shows savings as a function of time of work and commute time for other cities. NYC and Chicago have a similar distribution of savings like DC and the higher savings are observed for time of work between 10 am and 4 pm. In Boston, though the magnitude of savings is low generally, two bands of higher savings are observed for time of work between 1 am and 5 am, and between 11 am and 4 pm. The LBMP distribution in SFC is much like Phoenix, higher at night than in the day. Consequently, the savings in SFC are highest between 6 pm and 1 am. This shows how important LBMP and time of work are in determining V2G savings.

\subsection{Effect of Battery Efficiency and Charging Rates on V2G Savings}
\begin{figure}
    \centering
    \includegraphics[width=\textwidth]{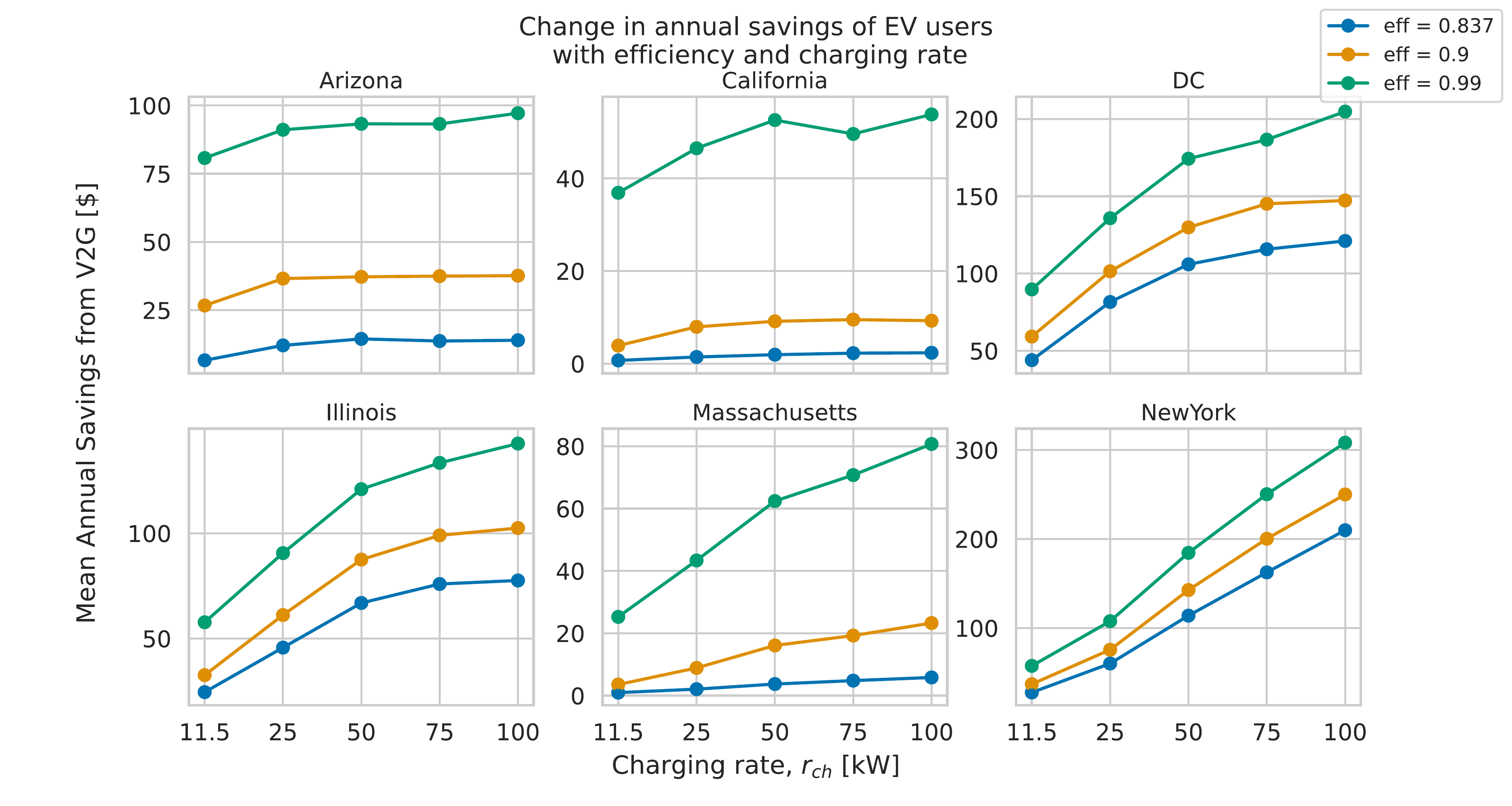}
    \caption{Effect of the battery efficiency and varying charging rates on the annual savings from V2G. It is observed that V2G savings increase with increase in one-way efficiency $\eta$, and charging rate $r_{ch}$. However, the rate of increase is different in different cities.}
    \label{fig:eff_effect}
\end{figure}
Charging efficiency is shown to have an effect on charging and discharging power losses\cite{Apostolaki-Iosifidou2017, Eftekhari2017, Li2015}. Studies are aimed at improving efficiency of lithium-ion batteries and finding new energy efficient materials \cite{Few2018}. The current one-way efficiency of lithium-ion batteries is 83.7\% \cite{Apostolaki-Iosifidou2018}. V2G analysis for higher efficiencies of 90\% and 99\% was done and compared to the current 83.7\% to see how the increasing efficiency would affect the annual V2G savings. Figure~\ref{fig:eff_effect} shows the results for this analysis. It is observed that as efficiency increases, V2G savings increase as the power losses during charging and discharging are cut down. 
Few studies have also demonstrated the effect of varying charging rates on profits to EV owners. \citet{Andersson2010} studied the effect of increasing charging rates on V2G (using PHEVs) for providing regulatory power in Sweden and Germany. They observed a four fold increase in PHEV users' profits when the charging rate was increased from 3.3kW to 15kW. Hence, the savings obtained at different efficiencies are also compared at various charging rates and higher charging rates result in increased savings from V2G. It can be seen in Figure~\ref{fig:eff_effect} that increasing the charging rate results in an increase in savings which is consistent with findings of \citet{Andersson2010}

\subsection{Future Battery Capital Costs and V2G}
\begin{figure}
    \centering
    \includegraphics[width=\textwidth]{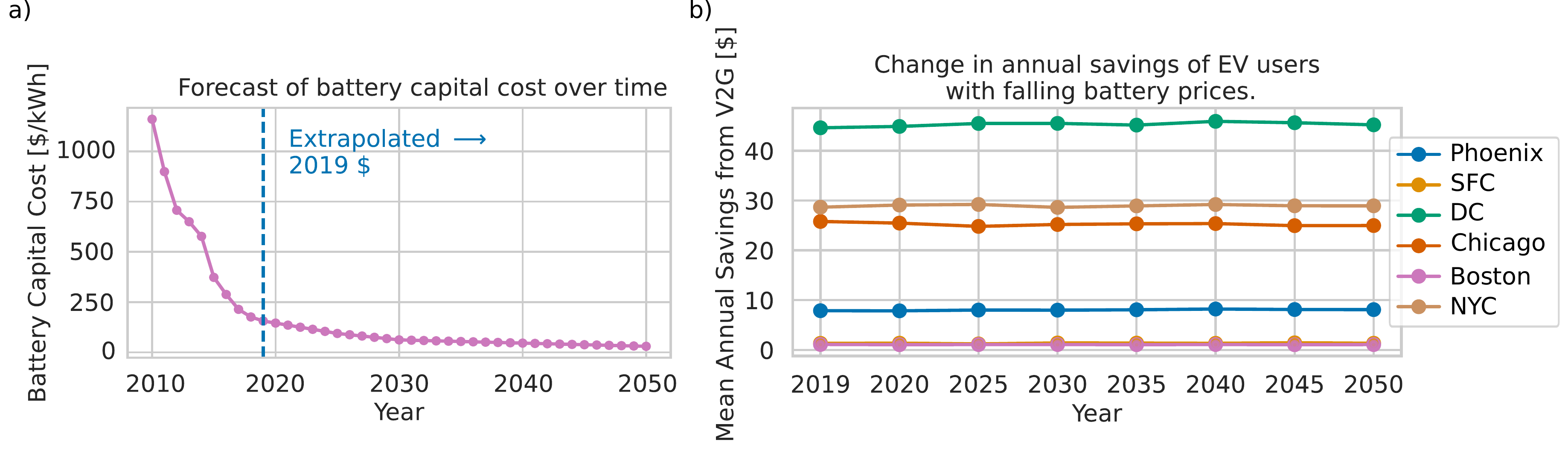}
    \caption{Panel (a): Lithium-ion battery cost extrapolated up to 2050. Battery capital cost data for 2010-2019 was taken from \citet{Bloomberg} report and it was extrapolated using exponential smoothing for the years 2020-2050. Panel (b): Change in mean annual savings for V2G users with changing battery capital cost over the years. The mean annual savings are calculated for V2G at the OSP which is optimized for every year. There is little effect of falling battery capital cost on the annual savings from V2G since the magnitude of degradation is very low.}
    \label{fig:future}
\end{figure}

The cost of lithium-ion batteries, which are most commonly used in EVs, has been falling and is projected to fall further from the current \$156/kWh to \$94/kWh by 2025 and to \$62/kWh by 2030 (Figure~\ref{fig:future}a).\cite{Bloomberg, citi, McKinsey} Will the falling price of batteries have any impact on savings earned from V2G at the current electricity prices? To answer this question, lithium-ion battery costs are extrapolated using exponential fitting up to 2050 and V2G economics calculated for every year for all cities previously mentioned. Current LBMPs and driving patterns are kept constant over the years and only battery capital cost is varied for these calculations. The OSP is optimized for every year.
Figure~\ref{fig:future} shows how the mean savings change with the change in the battery capital cost over the years. Savings remain almost constant for all cities as the battery prices fall indicating that battery cost has little effect. 

\subsection{Comparison with related work}
The results of this paper were compared with similar studies and conclusions were found to be similar in some cases. In the strategies proposed by \citet{Wang2016, Ma2012, Colmenar-Santos2017}, they observe that V2G effectively reduces mobility fuel cost if EV charging and discharging is done smartly. \citet{Colmenar-Santos2017} use `dumb-charging' versus `smart-charging' scenarios to calculate benefits from vehicle-to-home in Canary islands and find that the annual savings of users compared to just commute are $\pounds84$ ($\sim$\$103) in the `dumb-charging' case and $\pounds135$ ($\sim$\$165) in the `smart-charging' case. The dumb- and smart-charging cases used by them are analogous to the price-taker and OSP scenario in this paper but the scenarios in \citet{Colmenar-Santos2017} consider that the transactions are not happening over the meter. The techno-economic study by \citet{Dufo-Lopez2015} concluded that V2G storage system is not profitable for consumers in Spain at the current battery costs. They conclude that the battery capital cost must decrease and battery efficiency must improve before V2G is profitable to EV users in Spain. In contrast to these findings, \citet{GOUGH2017} claim that using V2G in the UK, if used for capacity markets, will make savings as high as $\pounds 8400$ ($\sim$\$10300) in a 10 year period and these savings are said to account for the infrastructure cost. These large savings are possible in the UK due to the annual triads. This verifies the finding that profits from V2G are location dependent. In the current study, there are some EV users who save around \$120/year -- \$150/year which is close to the findings of \citet{GOUGH2017}

\section{Conclusions}
A stochastic model with a realistic degradation model, and empirical driving and work patterns was built to assess effects of battery performance on EV users' profits from V2G. EV charging rate plays a crucial role in V2G profitability from a microeconomic point of view. The effect of efficiency, charging rates, and battery capital cost on V2G savings was studied. Two economic scenarios were evaluated - price-taker scenario and an optimal selling price scenario. Giving users a choice to set a selling price is crucial to make V2G a profitable scheme. Savings from V2G are highly location dependent due to a difference in electricity pricing. We find that an increasing charging rate and efficiency are much more important than the capital cost of lithium-ion batteries across all cities. Therefore, focusing future research efforts toward increasing EV charging rates and efficiency is more important than increasing EV battery capacity for V2G or cycle lifetime.

\section{Acknowledgement}
Authors would like to thank Center for Integrated Research Computing (CIRC) for providing computational resources and technical support. This research did not receive any specific grant from funding agencies in the public, commercial, or not-for-profit sectors.


\bibliographystyle{unsrtnat}
\bibliography{v2g.bib}

\newpage
\begin{center}
\Large \textsc{ Supplemental Information: City-wide modeling of Vehicle-to-Grid Economics to Understand Effects of Battery Performance}
\end{center}
\setcounter{section}{0}
\setcounter{equation}{0}
\setcounter{figure}{0}
\setcounter{table}{0}
\setcounter{page}{1}
\makeatletter
\renewcommand{\thesection}{S\arabic{section}}
\renewcommand{\theequation}{S\arabic{equation}}
\renewcommand{\thefigure}{S\arabic{figure}}
\renewcommand{\bibnumfmt}[1]{[S#1]}
\renewcommand{\citenumfont}[1]{S#1}

\section{Savings and work patterns}

\begin{figure}[!htb]
    \centering
    \includegraphics[width=0.9\textwidth]{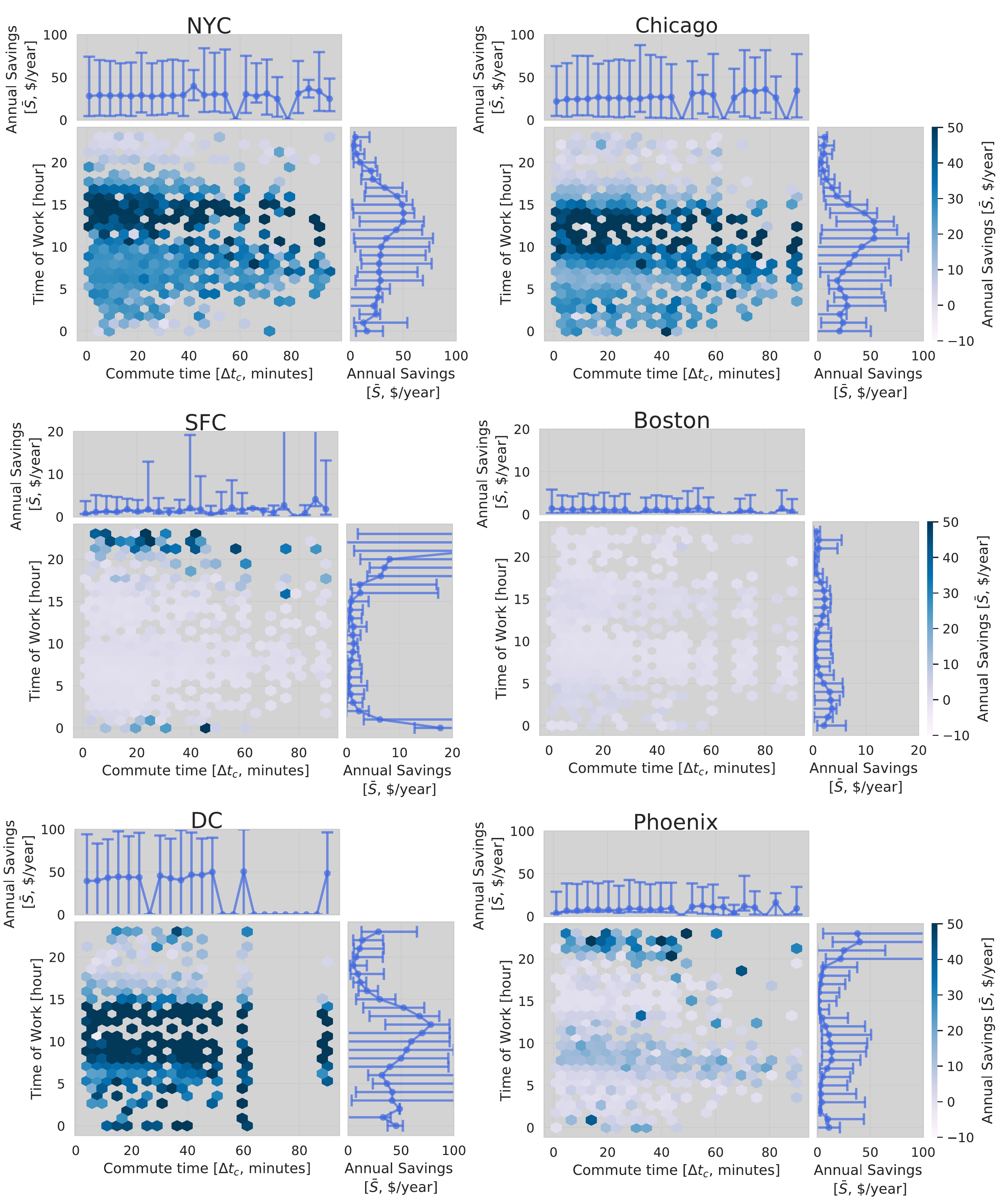}
    \caption{Variation in annual savings with commute time and time of work for the OSP scenario.}
    \label{fig:time_all}
\end{figure}

\begin{figure}[!htb]
    \centering
    \includegraphics[width=0.65\textwidth]{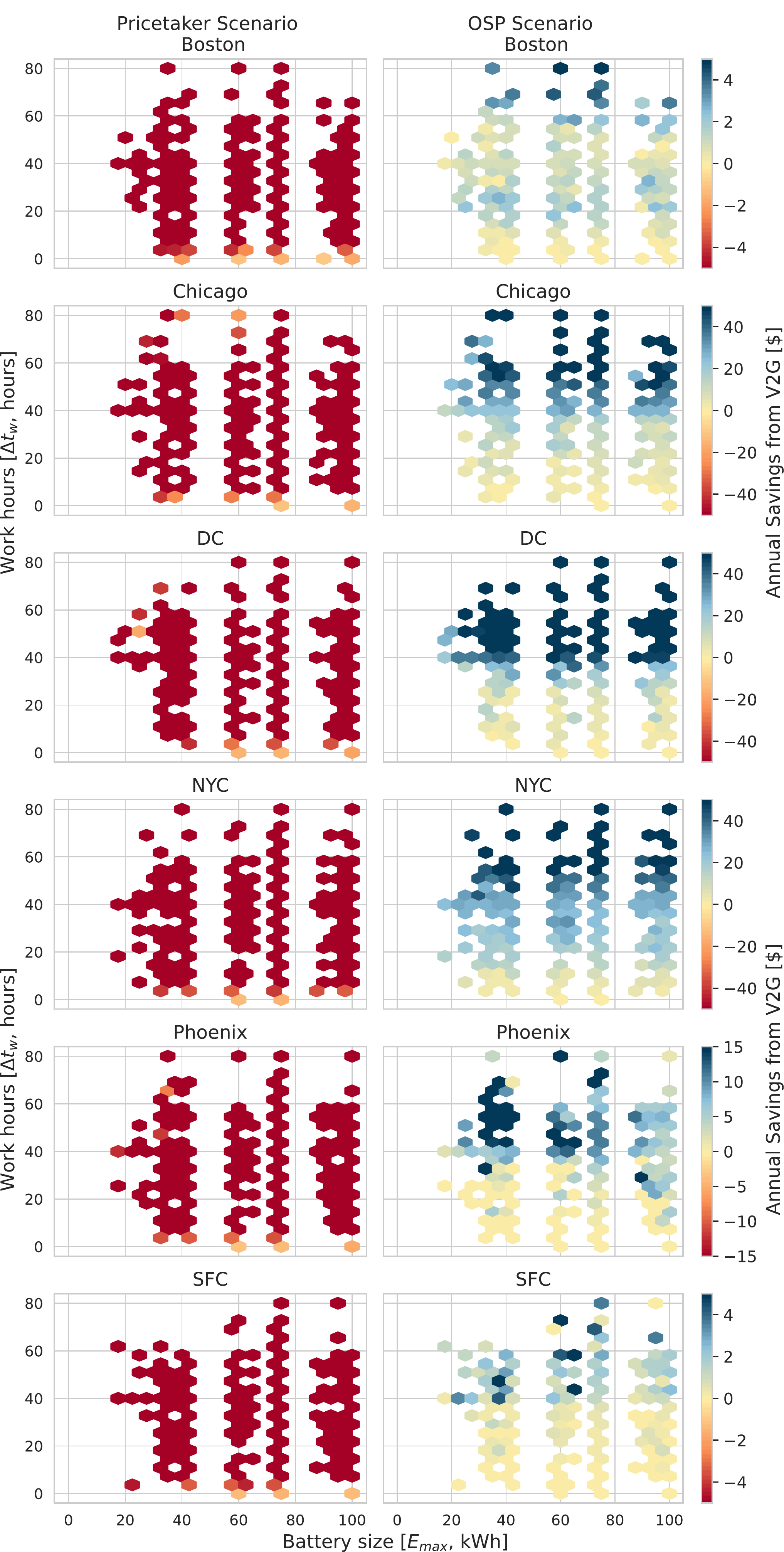}
    \caption{Annual savings from V2G in the Price-taker and OSP scenario as a function of battery size and average number of hours worked per week}
    \label{fig:battery_workhours_all}
\end{figure}

\end{document}